\documentclass[twocolumn,aps,prd,superscriptaddress,nofootinbib,showpacs
]{revtex4-1}

\usepackage{graphicx}
\usepackage{subcaption}
\usepackage{color}
\graphicspath{{figures/}{fig/}}
\usepackage{amsmath}
\usepackage{amssymb}
\usepackage{bm}
\usepackage{slashed}
\usepackage{amsfonts}
\usepackage{epstopdf}
\usepackage{bbm}
\usepackage{textcomp}
\usepackage{color}

\newcommand{\sect}[1]{\section{#1}}
\newcommand{\mi}{\mathrm{i}}
\newcommand{\md}{\mathrm{d}}

\begin{document}
\title{Complete two-loop electroweak corrections to $e^+e^-\rightarrow HZ$ }

\author{Xiang Chen}
\email{xchenphy@pku.edu.cn}
\affiliation{School of Physics, Peking University, Beijing 100871, China}

\author{Xin Guan}
\email{guanxin0507@pku.edu.cn}
\affiliation{School of Physics, Peking University, Beijing 100871, China}

\author{Chuan-Qi He}
\email{legend\_he@pku.edu.cn}
\affiliation{School of Physics, Peking University, Beijing 100871, China}

\author{Zhao Li}
\email{zhaoli@ihep.ac.cn}
\affiliation{Institute of High Energy Physics, Chinese Academy of Sciences, Beijing 100049, China}
\affiliation{School of Physics Sciences, University of Chinese Academy of Sciences, Beijing 100039, China}
\affiliation{Center for High Energy Physics, Peking University, Beijing 100871, China}

\author{Xiao Liu}
\email{xiao.liu@physics.ox.ac.uk}
\affiliation{Rudolf Peierls Centre for Theoretical Physics, Clarendon Laboratory, Parks Road, Oxford OX1 3PU, UK}

\author{Yan-Qing Ma}
\email{yqma@pku.edu.cn}
\affiliation{School of Physics, Peking University, Beijing 100871, China}
\affiliation{Center for High Energy Physics, Peking University, Beijing 100871, China}

\begin{abstract}
We compute the complete two-loop electroweak corrections to the Higgsstralung process $e^+e^-\rightarrow HZ$ at the future Higgs factory. The Feynman integrals involved in the computation are decomposed into linear combinations of a minimal set of master integrals taking advantage of the recent developments of integral reduction techniques. The master integrals are then evaluated by differential equations with boundary conditions provided by the auxiliary mass flow method. Our final result for given $\sqrt{s}$ is expressed as a piecewise function defined by several deeply expanded power series, which has high precision and can be further manipulated efficiently.  Our calculation presents the first complete two-loop electroweak corrections for processes with four external particles. 
\end{abstract}

\maketitle

\sect{Introduction}
The physics related to the Higgs boson has become the frontier of the high energy physics since its discovery a decade ago \cite{ATLAS:2012yve,CMS:2012qbp}. In the Standard Model (SM) of particle physics, the Higgs boson is known as the direct evidence of the electroweak (EW) spontaneous symmetry breaking based on the Higgs mechanism. 
However, the current experiment precision cannot exclude the possibility of exotic Higgs potential deviated from the SM, which is the typical structure in most of the new physics models.
Therefore, the Higgs boson could be the most promising probe to new physics beyond the SM. 

The precise measurement on the Higgs boson is the most important mission for the next generation of high-energy experiment facility. In the past few years, there have been several proposals of the Higgs factory, including  the International Linear Collider (ILC) \cite{Baer:2013cma,Behnke:2013xla,Bambade:2019fyw}, the Circular Electron Positron Collider (CEPC) \cite{CEPCStudyGroup:2018ghi, CEPCStudyGroup:2018rmc}, and the Future Circular Collider (FCC-ee) \cite{TLEPDesignStudyWorkingGroup:2013myl,FCC:2018byv,FCC:2018evy}. Millions of Higgs bosons are expected to be produced in these Higgs factories via the processes including the Higgsstrahlung $e^+e^-\rightarrow ZH$, the W boson fusion $e^+e^-\rightarrow \nu_e \bar \nu_e H$, and the Z boson fusion $e^+e^-\rightarrow e^+e^-H$. At a typical center-of-mass energy $240$~GeV, the dominant contribution is the Higgsstrahlung process. 

The preliminary investigations have shown that, with the expected integrated luminosity of 5.6 ab${}^{-1}$ \cite{An:2018dwb}, the Higgsstrahlung cross section $\sigma(e^+e^-\rightarrow ZH)$ can be measured with the precision $0.51\%$ \cite{An:2018dwb}. Consequently, the precision of the relevant theoretical predictions must be pushed to at least the same level. The next-to-leading order (NLO) EW effect has been investigated in Refs.\cite{Fleischer:1982af,Kniehl:1991hk,Denner:1992bc}, and in recent years the mixed EW-QCD correction effect has been obtained \cite{Gong:2016jys,Sun:2016bel,Chen:2018xau,Ma:2021cxg}. Even though, the current theoretical uncertainty is still as large as 1\%, which is not compatible to the experiment accuracy yet. Therefore, higher order radiative corrections are crucial for the Higgs physics analysis at the future Higgs factory.
Recently some integrals involved in two-loop EW corrections were calculated  \cite{Song:2021vru,Liu:2021wks}, but complete two-loop calculation is still missing.

The complete two-loop EW calculation for this $2\to 2$ process is always challenging but is indispensable for the reliable theoretical predictions. On one hand, 25377 Feynman diagrams~\cite{Li:2020ign} make all subsequent procedures very time and resource consuming. And on the other hand, the Feynman integrals involving six mass scales are out of the reach of all analytical toolkit. In this paper, by taking advantage of the recent developments of multiloop calculation techniques, we eventually achieve the complete two-loop EW corrections for the Higgsstralung process $e^+e^-\rightarrow HZ$ at the future Higgs factory.

The rest of the paper is organized as follows.
In Section 2, technique details of the calculation of two-loop EW correction are explained. In Section 3, the $\gamma_5$ scheme and renormalization are discussed. In Section 4, we present the numerical results of two-loop EW corrections. The summary is made in the last section.

\sect{Two-loop calculation}

We consider the two-loop electroweak correction to the following process
\begin{equation}
    \label{equ:process}
    e^+(k_1)+e^-(k_2)\rightarrow H(k_3)+Z(k_4),
\end{equation}
where the external momenta satisfy the on-shell conditions 
\begin{align}
k_1^2=k_2^2=0,\quad k_3^2=m_H^2,\quad k_4^2=m_Z^2,
\end{align}
and momentum conservation
\begin{align}
k_1+k_2=k_3+k_4.
\end{align}
The Mandelstam variables are defined as usual
\begin{align}
s=(k_1+k_2)^2,\quad t=(k_1-k_3)^2.
\end{align}

\begin{figure}[htb]
	\centering
    \begin{minipage}[b]{.451\linewidth}
        \centering
        \includegraphics[width=1.0\linewidth]{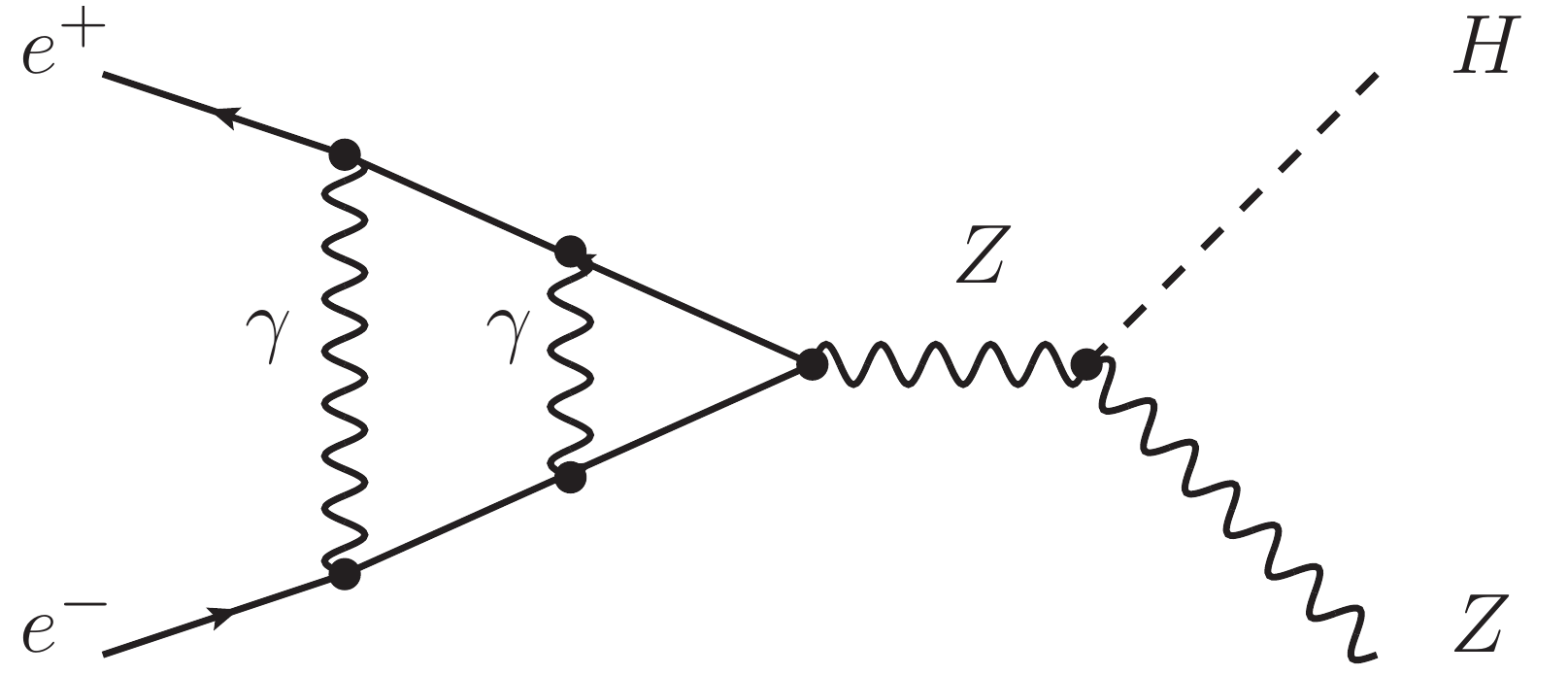}
        \subcaption{}
    \end{minipage}
    \begin{minipage}[b]{.451\linewidth}
        \centering
        \includegraphics[width=1.0\linewidth]{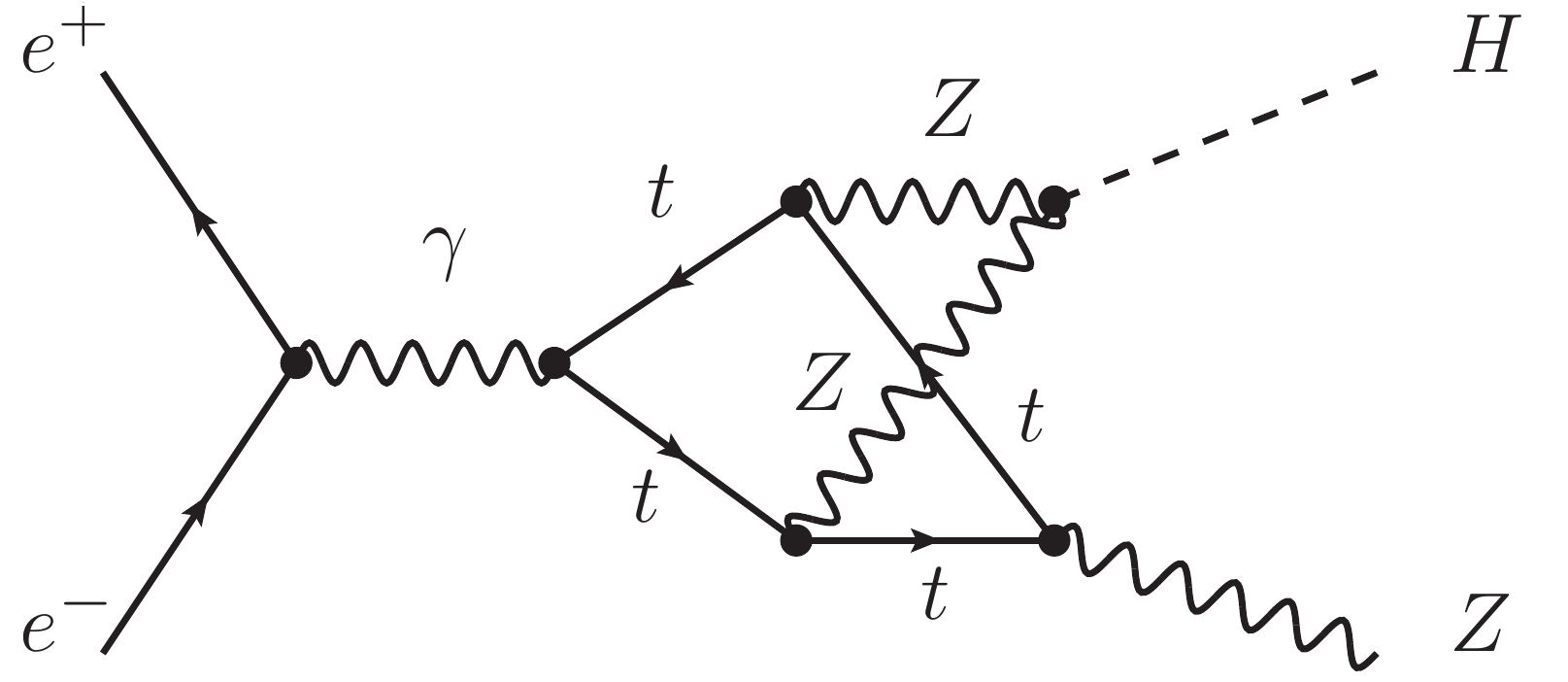}
        \subcaption{}
    \end{minipage}
    \begin{minipage}[b]{.451\linewidth}
        \centering
        \includegraphics[width=1.0\linewidth]{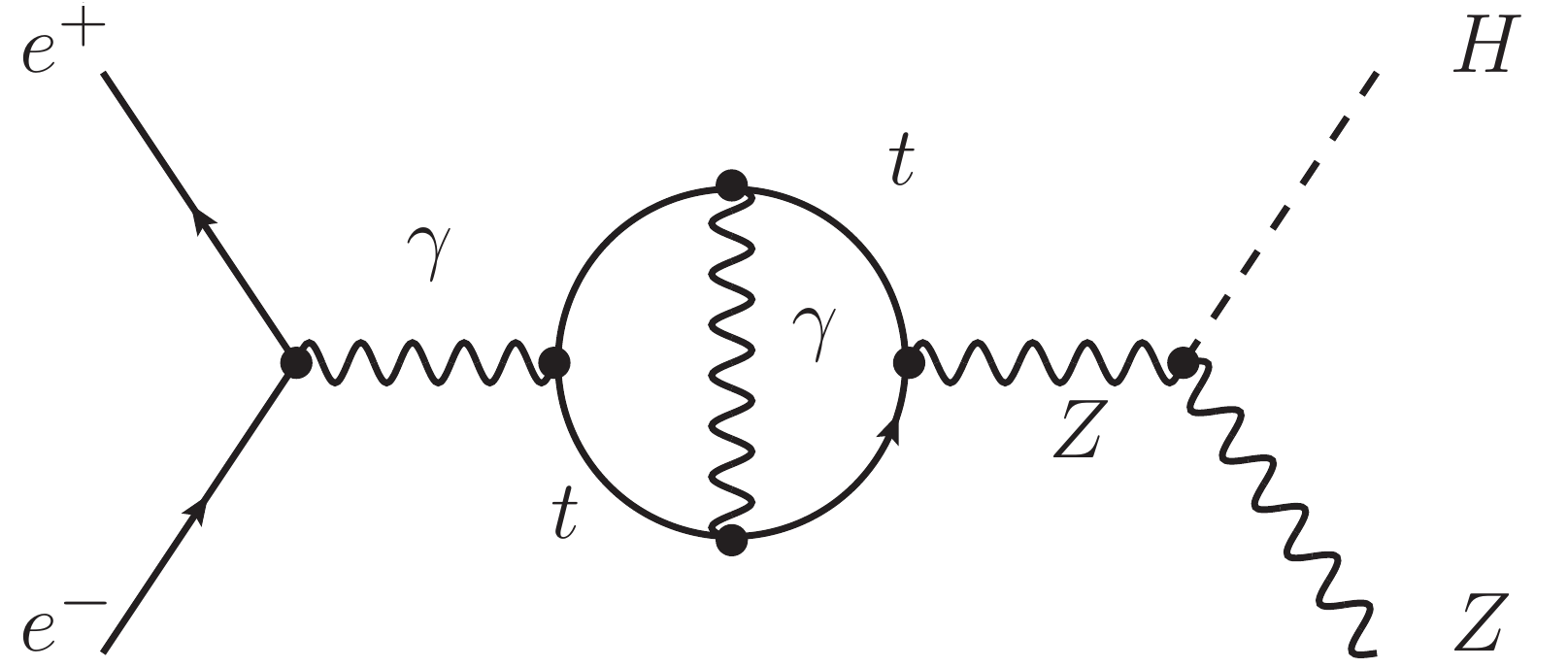}
        \subcaption{}
    \end{minipage}
    \begin{minipage}[b]{.451\linewidth}
        \centering
        \includegraphics[width=1.0\linewidth]{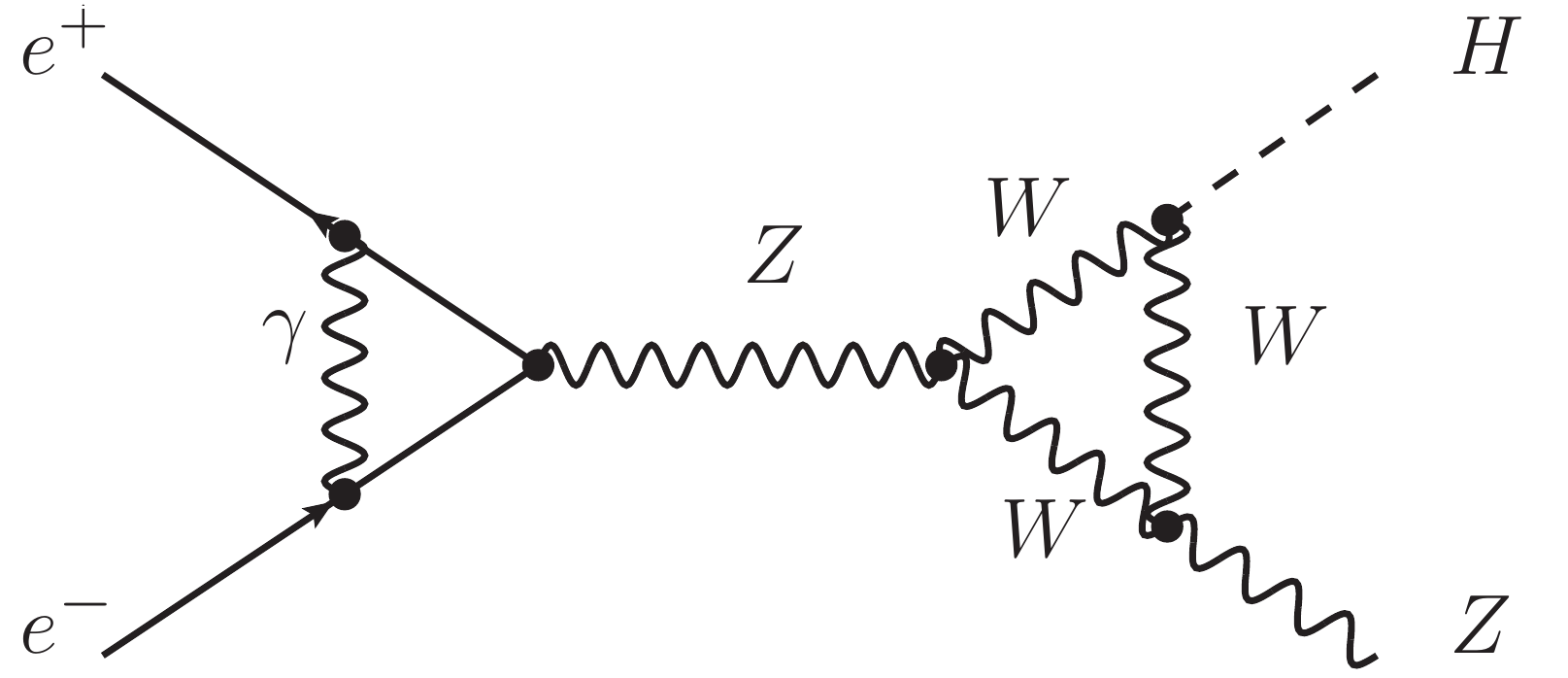}
        \subcaption{}
    \end{minipage}
    \begin{minipage}[b]{.451\linewidth}
        \centering
        \includegraphics[width=1.0\linewidth]{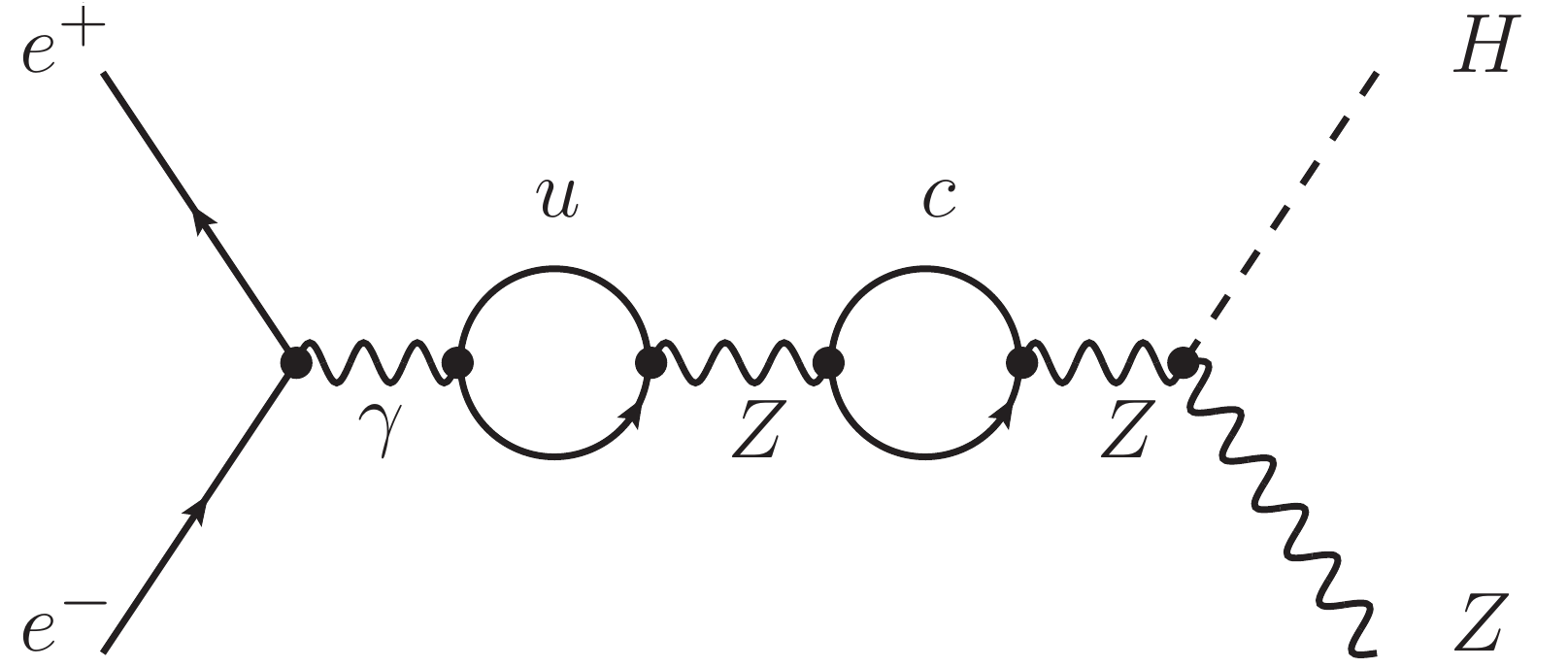}
        \subcaption{}
    \end{minipage}
    \begin{minipage}[b]{.451\linewidth}
        \centering
        \includegraphics[width=1.0\linewidth]{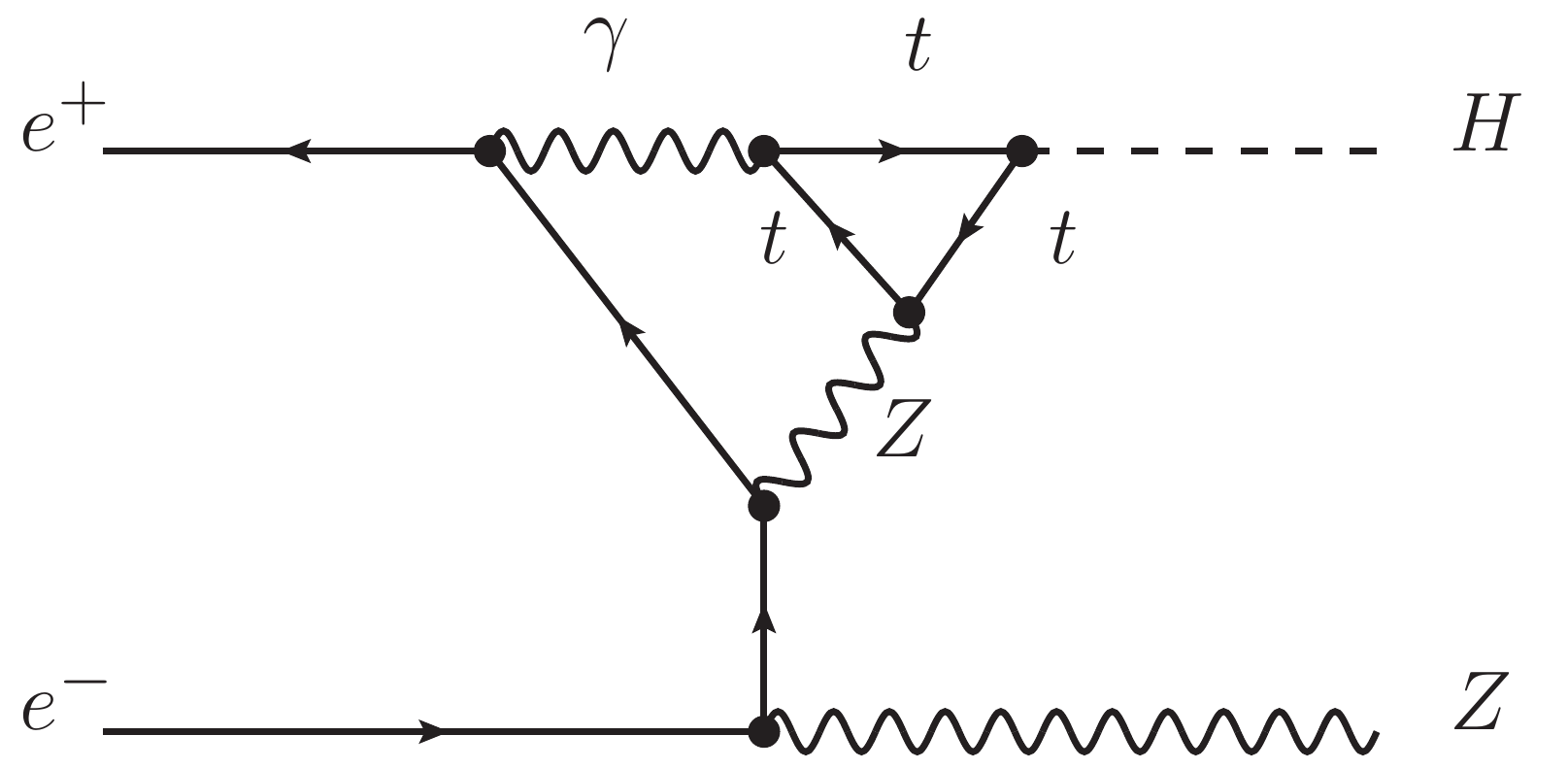}
        \subcaption{}
    \end{minipage}
    \begin{minipage}[b]{.451\linewidth}
        \centering
        \includegraphics[width=1.0\linewidth]{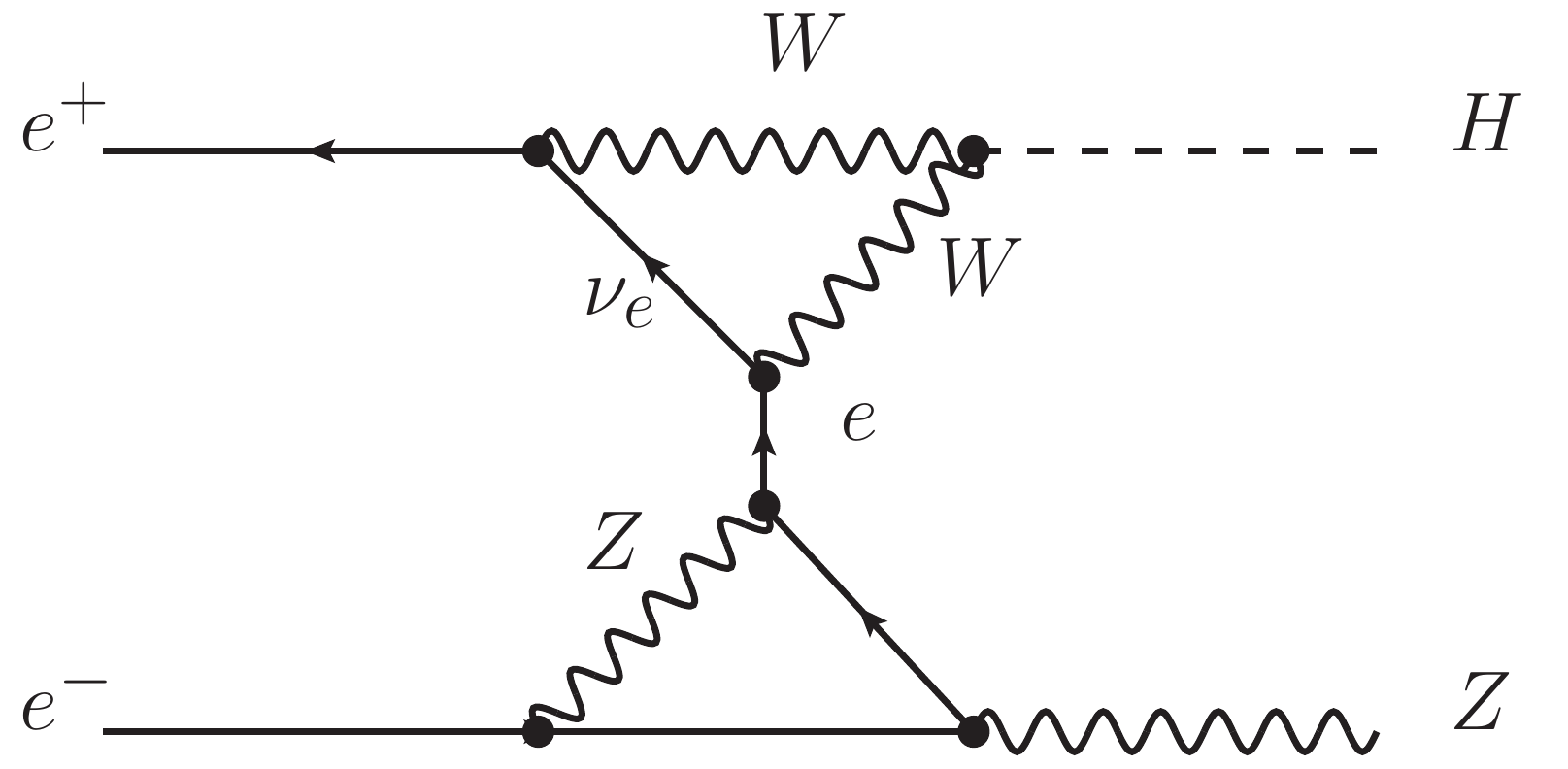}
        \subcaption{}
    \end{minipage}
    \begin{minipage}[b]{.451\linewidth}
        \centering
        \includegraphics[width=1.0\linewidth]{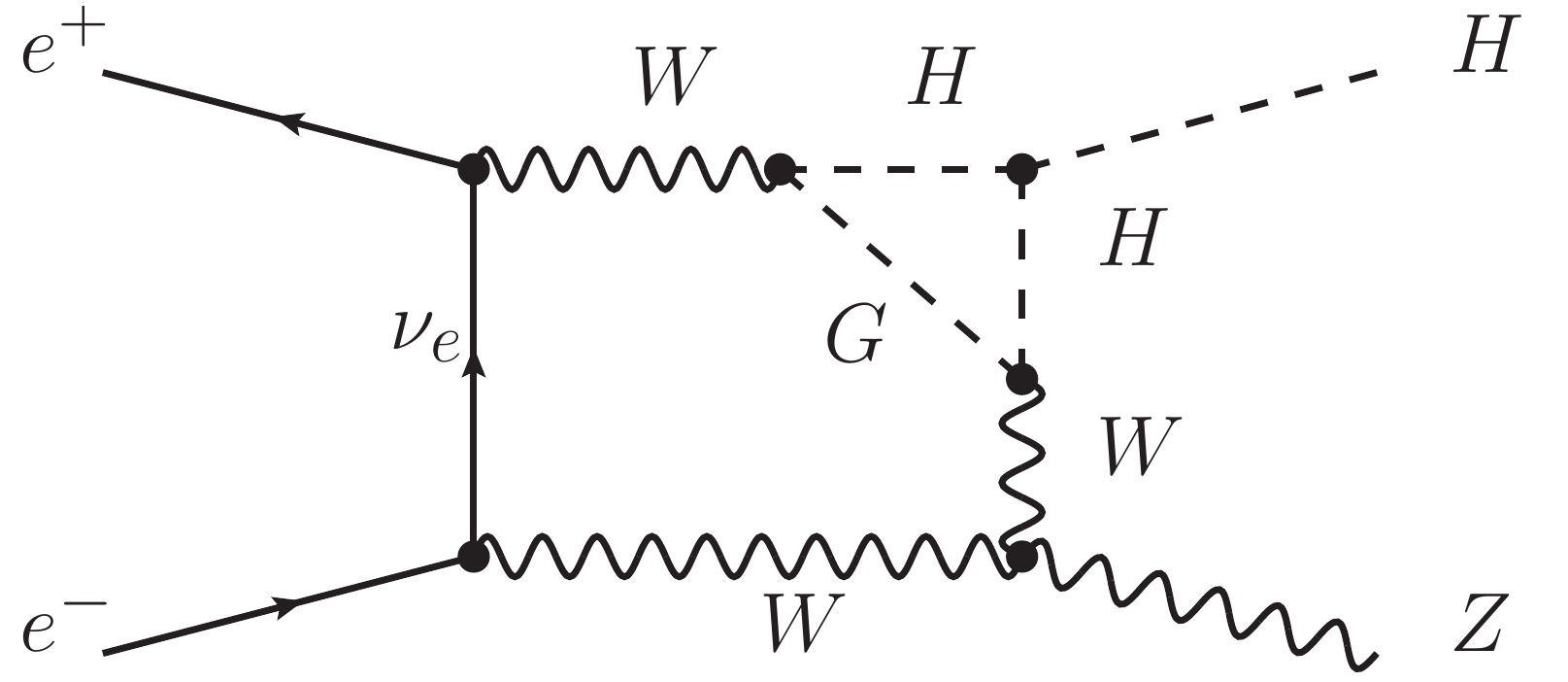}
        \subcaption{}
    \end{minipage}
    \begin{minipage}[b]{.451\linewidth}
        \centering
        \includegraphics[width=1.0\linewidth]{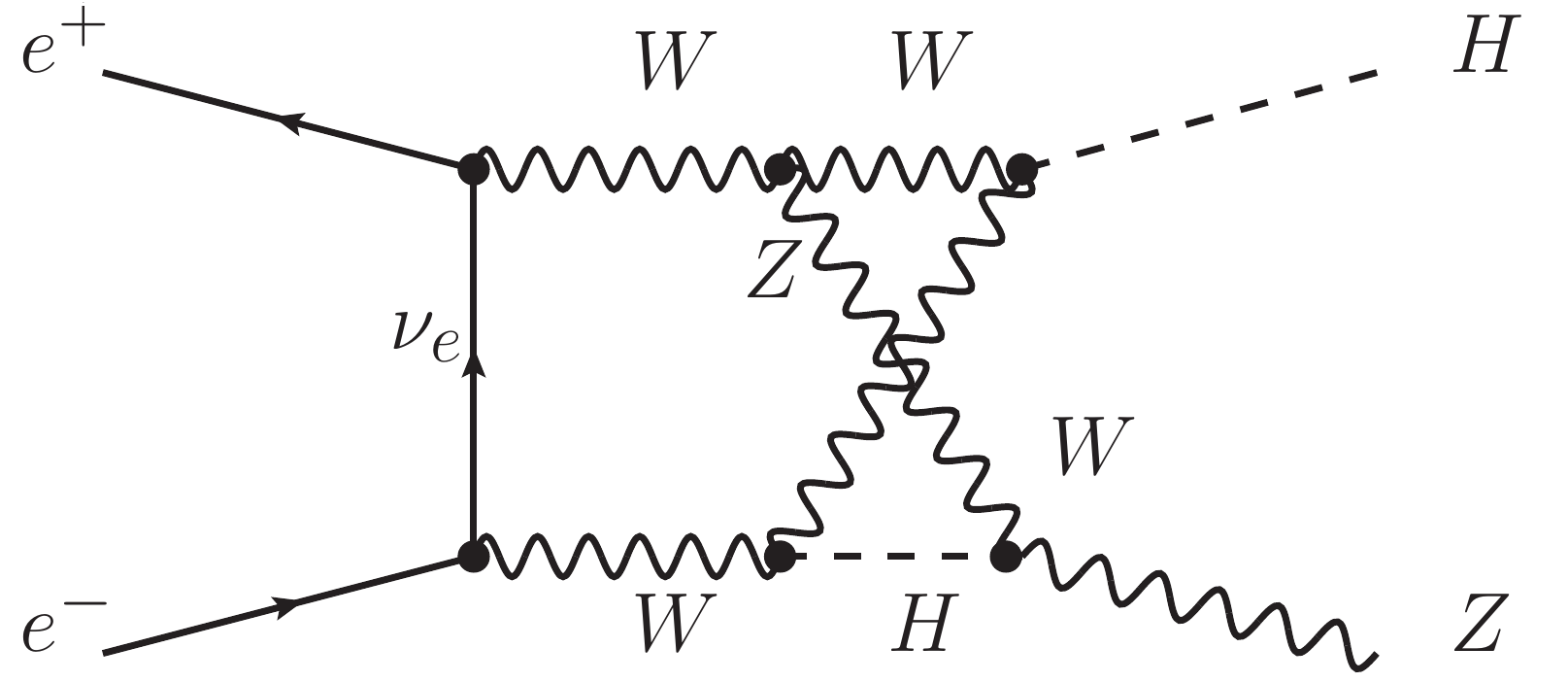}
        \subcaption{}
    \end{minipage}
	\caption{\label{fig:FeynmanDiagram}
			Some sample Feynman diagrams at the two-loop level. 
}
\end{figure}

We generate the Feynman amplitudes with {\tt QGRAF}~\cite{Nogueira:1991ex,Nogueira:2021wfp} and {\tt FeynArts}~\cite{Hahn:2000kx}, with some sample two-loop Feynman diagrams shown in Fig~\ref{fig:FeynmanDiagram}. A detailed classification of all 25377 two-loop diagrams have been provided in Ref~\cite{Li:2020ign}, from which one can reckon the complexity of the calculation.  We use a private {\tt Mathematica} package to deal with Lorentz algebras, and express the interference of two-loop amplitudes with tree-level amplitudes as linear combinations of some scalar integrals. The coefficients of these integrals are rational functions of physical parameters, including the Mandelstam variables $s$ and $t$, squared mass of corresponding particles $m_t^2$, $m_H^2$, $m_Z^2$ and $m_W^2$, dimensional regulator $\epsilon=(4-D)/2$, and the electroweak coupling constant $\alpha$.

Totally we get about $3\times 10^{4}$ Feynman integrals, which are then clustered into 372 integral families. To illustrate how to evaluate these integrals, let us take the integral family defined by the Feynman diagram shown in Fig.~\ref{fig:FeynmanDiagram} (i) as an example, which is one of the most complicated Feynman integral family. The integrals in this family can be expressed as
\begin{align}\label{eq:ints}
I(\nu_1,\cdots,\nu_9)=\int\prod_{i=1}^{L}\frac{\md^{D}\ell_i}{\mi\pi^{D/2}}
\frac{\mathcal{D}_{8}^{-\nu_{8}}\mathcal{D}_9^{-\nu_9}}{\mathcal{D}_1^{\nu_{1}}\cdots \mathcal{D}_7^{\nu_{7}}},
\end{align}
where inverse propagators can be chosen as
\begin{align}\label{eq:propgator}
&\mathcal{D}_1 = \ell_1^2, \, \mathcal{D}_2=(\ell_1+k_1)^2-m_W^2,\, \mathcal{D}_3=(\ell_1-k_2)^2-m_W^2,\nonumber\\
&\mathcal{D}_4=(\ell_1-\ell_2)^2-m_H^2,\,\mathcal{D}_5=(\ell_1-\ell_2+k_4)^2-m_Z^2,\nonumber\\
&\mathcal{D}_6=(\ell_2-k_2)^2-m_W^2,\,\mathcal{D}_7=(\ell_2+k_1-k_4)^2-m_W^2,\nonumber\\
&\mathcal{D}_8 = (\ell_1-\ell_2+k_1)^2,\,\mathcal{D}_9=(\ell_1-\ell_2+k_2)^2,
\end{align}
with the last two being irreducible scalar products introduced for completeness. For simplicity, during the computation, we fix masses of particles as rational numbers. Then for any given rational value of $s$, the integrals only depend on two variables, $t$ and $\epsilon$.

We first decompose our target integrals into linear combinations of a smaller set of so-called master integrals using integration-by-parts (IBP)~\cite{Chetyrkin:1981qh} reduction\footnote{Integrals needed for constructing differential equations are also included in the target integrals, see next paragraph.}. In detail, we first use {\tt LiteRed}~\cite{Lee:2013mka} and {\tt FiniteFlow}~\cite{Peraro:2019svx} to generate and solve the system of IBP identities~\cite{Chetyrkin:1981qh} based on Laporta's algorithm~\cite{Laporta:2000dsw} over finite field. Around 200 numerical samplings are sufficient to construct the block-triangular relations  for target integrals proposed in Refs. ~\cite{Liu:2018dmc, Guan:2019bcx}. We then make full use of the block-triangular relations to efficiently generate large amount of samplings (approximately $10^4$) to eventually reconstruct the reduction coefficients. This strategy reduces the computational time by several times, compared with the reduction without using block-triangular relations.

Next we compute the master integrals using differential equations~\cite{Kotikov:1990kg} based on power series expansion~\cite{Caffo:2008aw, Czakon:2008zk}. The differential equations of master integrals with respect to $t$ are constructed using aforementioned IBP reduction. The boundary conditions, say at $t/m_t^2=-1/2$, are then fixed by the auxiliary mass flow method~\cite{Liu:2017jxz,Liu:2020kpc, Liu:2021wks, Liu:2022mfb} implemented in {\tt AMFlow}~\cite{Liu:2022chg}. More specifically, we employ the ``mass'' mode~\cite{Liu:2021wks}, to insert the auxiliary mass parameter $\eta$ to propagators $\mathcal{D}_2, \mathcal{D}_3, \mathcal{D}_6$ and $\mathcal{D}_7$. This is equivalent to directly treat $m_W^2$ as a dynamical parameter. By doing so, we get the integrals simplified, in the large mass limit~\cite{Beneke:1997zp, Smirnov:1999bza}, to some factorized one loop integrals and vacuum integrals. These simplified integrals will be then evaluated again by using the {\tt AMFlow}. Finally, a numerical continuation of the auxiliary mass from the large mass limit to zero would give us high precision physical results, which serves as the boundary conditions. With these in hand, we are able to construct a piecewise function for each master integral represented by some deeply expanded power series expansions, which is straightforward using the differential equations solver in {\tt AMFlow}. During the numerical evaluations, $\epsilon$ are set to some fixed values and the $\epsilon$ dependence is only reconstructed at the final stage as proposed in Ref.~\cite{Liu:2022mfb,Liu:2022chg}. In this way, we do not need to manipulate Laurent expansions of $\epsilon$ during the intermediate stages of calculations, which significantly reduces the computational time.

\begin{figure}[htb]
	\begin{center}
		\includegraphics[width=0.9\linewidth]{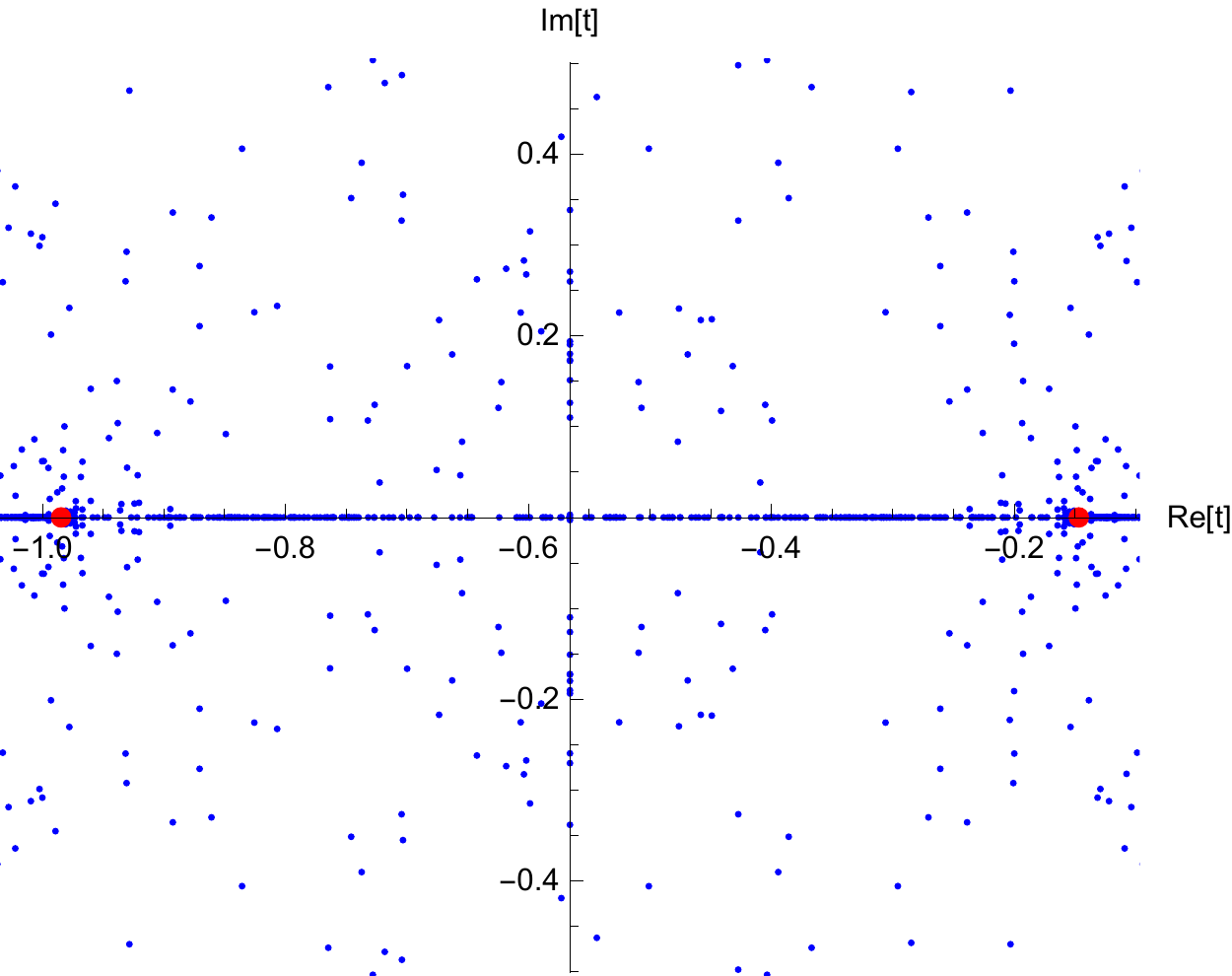}
		\caption{\label{fig:SingularityDistribution}
			Singularities distribution. Blue dots are singularities read from differential equations. Two red dots on the real axis are the boundary points of physical region.
 }
	\end{center}
\end{figure}

One of the difficulties to solve the differential equations is that there are many singularities, shown in Fig~\ref{fig:SingularityDistribution}, as a result of multiscaleness of this process. By investigating asymptotic behaviours near each singularity, with the help of differential equations and boundary conditions, we find only a few of them are physical, while all the others are non-physical, arising probably from the ``bad'' choice of master integrals or singularities in other Riemann sheets. Although most singularities are non-physical, they would affect the stability of our numerical computations. Therefore we must carefully design a segmentation of the physical region to reduce these effects. By trial and error, we finally obtain a possible segmentation, where the physical region are divided into more than 20 pieces. 
We perform a deep power series expansion in each piece such that, after reconstructing the $\epsilon$ dependence, high precision are obtained in the whole physical region.

The evaluation of integrals in other families are quite similar. To summarize, we get 7675 master integrals in total after reduction. The integrals reduction along with the differential equations construction cost about $\mathcal{O} (10^4)$ CPU $\cdot$ h. Calculating the boundary conditions at $t= -1/2$ cost about $1.5 \times 10^4$ CPU $\cdot$ h. And the piecewise functions of master integrals are obtained in $3 \times 10^3$ CPU $\cdot$ h.

\sect{$\gamma_5$ scheme and Renormalization}
It is well known that, in dimensional regularization, the anticommutation relation $\{\gamma^\mu, \gamma_5\}$ and the cyclicity of Dirac trace cannot be satisfied simultaneously.  
In practice, we prefer to keep the anticommutation relation $\{\gamma^\mu, \gamma_5\}$ not only to simplify the computation, but also because using a non-anticommute $\gamma_5$ with the other Dirac matrices $\gamma^\mu$ leads to ``spurious anomalies" which violate chiral symmetry and hence gauge invariance~\cite{Jegerlehner:2000dz}. In this case, one must impose the relevant Ward-Takahashi (WT) and Slavnov-Taylor identities order by order to keep the renormalizability of a perturbation theory, which will make practical calculations much more difficult and tedious obviously. Thus we choose the so-called na\"{i}ve-$\gamma_5$ scheme so that anticommutation relations hold. Then a new problem arises: where should we read a fermion loop? Due to the lack of the cyclicity of the trace, reading a fermion loop from a different point may give different final result. This ambiguity can be fixed by  the KKS-scheme~\cite{Korner:1991sx}, where the final result is defined as the average of all possible reading points starting from a  $\gamma_5$.

Once the $\gamma_5$ scheme is fixed, one can study the renormalization of ultraviolet divergences. For two-loop EW corrections, this includes  renormalization of masses, couplings, wave functions, field mixings, and so on. In literature,  frequently used renormalization schemes include on-shell scheme, such as in Ref.~\cite{Freitas_2000}, and $\overline{\mathrm{MS}}$ scheme. Results obtained in different schemes can be related by a finite renormalization, and thus are equivalent.

\sect{Numerical Results}
By combining everything together, we finally obtain the result of two-loop electroweak contribution $\mathcal{A}^{(2)}$ defined by
\begin{align}
    \mathcal{A}^{(2)}=2\sum_{\text{spin}}\Re\text{e}\left(\mathcal{M}^{(2)}\mathcal{M}^{(0)*}\right),
\end{align}
where $\mathcal{M}^{(2)}$ is the two-loop amplitude and $\mathcal{M}^{(0)}$ is the tree level amplitude.
Our computation is carried out under the following numerical configuration
\begin{equation}
{m_H^2\over m_t^2}={12\over 23},\quad
{m_Z^2\over m_t^2}={23\over 83},\quad
{m_W^2\over m_t^2}={14\over 65},
\end{equation}
which correspond to $m_t=172.69\text{ GeV}$~\cite{Workman:2022ynf}, $m_H=124.7\text{ GeV}$, $m_Z=90.906\text{ GeV}$ and $m_W=80.145\text{ GeV}$.

As a benchmark, we provide numerical results with the following phase space point
\begin{equation}
{s\over m_t^2}={83\over 43},
\quad
{t\over m_t^2} = -{5\over 7},
\end{equation}
or $\sqrt{s}=240\text{ GeV}$ and $t=-21301.3\text{ GeV}^2$.
Then we get
\begin{align}
    \mathcal{A}^{(2)}=\alpha^4(&75548.083\epsilon^{-4}\notag\\
    -&3.1962821\times 10^{6}\epsilon^{-3}\notag\\
    +&1.1548893\times 10^{7}\epsilon^{-2}\notag\\
    +&2.6990603\times 10^{8}\epsilon^{-1}\notag\\
    +&1.5608903\times 10^{9}+\mathcal{O}(\epsilon)),
\end{align}
where divergences should be canceled if we sum over all contributions at NNLO level, including real emissions.
In addition, we provide the result of the corner integral of the top sector in Eq.~\eqref{eq:ints}
\begin{align}
    I&(1,1,1,1,1,1,1,0,0)=\notag\\
    &(0.35203833-16.253246 \mi)\notag\\
    &+(10.998841-64.231845 \mi)\epsilon\notag\\
    &+(32.180275-134.31458 \mi)\epsilon^{2}\notag\\
    &+(45.366882-198.45944 \mi)\epsilon^{3}\notag\\
    &+(27.957706-234.39361 \mi)\epsilon^{4}+\mathcal{O}(\epsilon^5).
\end{align}
For other values of $\sqrt{s}$ and $t$, results can be obtained similarly.

\sect{Summary}
In this paper we present the first complete calculation of two-loop EW corrections for Higgsstralung process $e^+e^-\rightarrow HZ$ at the future Higgs factory. Our result for given $\sqrt{s}$ is expressed as a piece-wise function defined by several deeply expanded power series. The results have high precision and can be use in future efficiently. This work is possible thanks to many state-of-the-art techniques. Our calculation represents the first complete two-loop electroweak corrections for processes with four external particles.

What we have done is the most difficult part of the complete NNLO EW corrections for $H+Z$ production at the future Higgs factory. In the near future, we will calculate also real-emission corrections and then the complete NNLO EW corrections can be achieved.

\section*{Acknowledgments}
We would like to thank Ayres Freitas, Yu Jia, Li Lin Yang and Yang Zhang for many helpful discussions.
This work is supported in part by the National Natural Science Foundation of
China (Grants  No. 11875071, No. 11975029, No. 12075251), the National Key Research and Development Program of China under
Contracts No. 2020YFA0406400, and the High-performance Computing Platform of Peking University. The research of XL was also supported by the ERC Starting Grant 804394 \textsc{HipQCD} and by the UK Science and Technology Facilities Council (STFC) under grant ST/T000864/1.
{\tt JaxoDraw}~\cite{BINOSI200476} was used to generate Feynman diagrams.

{\bf Note added:} When our paper was being finalized, a preprint appeared~\cite{Freitas:2022hyp} which presents the fermion-loop contributions to the same process.

\providecommand{\href}[2]{#2}\begingroup\raggedright\endgroup

\end{document}